\documentclass[journal]{IEEEtran}

\ifCLASSINFOpdf
\else
   \usepackage[dvips]{graphicx}
\fi
\usepackage{url}

\usepackage[T1]{fontenc}
\usepackage{graphicx}
\usepackage{multirow}
\usepackage{amsmath}
\usepackage{balance}
\usepackage[dvipsnames]{xcolor}
\begin{document}

\title{SSIM-Variation-Based Complexity Optimization for Versatile Video Coding}

\author{Jielian Lin,  Hongbin Lin, Zhichen Zhang, Yiwen Xu, \IEEEmembership{Member, IEEE} and Tiesong Zhao, \IEEEmembership{Senior Member, IEEE}

\thanks{This work is supported by the National Natural Science Foundation of China under Grants 62171134, and the Natural Science Foundation of Fujian Province under Grants 2019J01222. (*Corresponding author: Yiwen Xu, xu\_yiwen@fzu.edu.cn)}

\thanks{	J. Lin, H. Lin, Z. Zhang, Y. Xu and T. Zhao are with Fujian Key Lab for Intelligent
	Processing and Wireless Transmission of Media Information, Fuzhou
	University, Fuzhou, Fujian 350116, China. (e-mail: \{N191110005, 211127036, 211120117, xu\_yiwen, t.zhao\}@fzu.edu.cn). }
}

\markboth{Journal of \LaTeX\ Class Files, Vol. 14, No. 8, August 2015}
{Shell \MakeLowercase{\textit{et al.}}: Bare Demo of IEEEtran.cls for IEEE Journals}
\maketitle

\begin{abstract}
To date, Versatile Video Coding (VVC) has  a more magnificent overall performance than High Efficiency Video Coding (HEVC). The Quadtree with Nested Multi-Type Tree (QTMT) coding block structure  can substantially   enhance video coding quality in VVC. However, the coding gain also leads to a greater coding complexity. Therefore,  this letter proposes a Fast Decision Scheme Based on Structural Similarity Index Metric Variation (FDS-SSIMV)  to solve this problem. Firstly, the  Structural Similarity Index Metric Variation (SSIMV) characteristic among the sub coding units of the spit mode is illustrated. Next, to  evaluate the SSIMV value,  SSIMV measure strategies are designed  for different split modes in this letter. Then, the desired split modes are selected by the SSIMV values. Experimental results show that the proposed method achieves  64.74\%  average encoding Time Saving (TS) with a 2.79\% Bjøntegaard Delta Bit Rate (BDBR), outperforming  the benchmarks.
\end{abstract}


\IEEEpeerreviewmaketitle

\section{Introduction}
\label{sec:intro}

With the development of video coding technology, the up-to-date video coding standard has been updated to Versatile Video Coding (VVC) \cite{introduceVVC} by the Joint Video Exploration Team (JVET). The goal of VVC  is to reduce the bit rate by 50\% than High Efficiency Video Coding (HEVC) \cite{introduceH265} at the same quality. To achieve this goal, many new coding techniques have been explored in the newest standard. However, these techniques also lead to a significant increase in the complexity of video coding. Therefore, the  complexity optimization of VVC has become one of the most concerned issues of researchers.

\begin{figure}[tbp]
	
	\centering
	\begin {minipage} [htbp] {0.42\textwidth}
	\centering
	\includegraphics [width=1\textwidth] {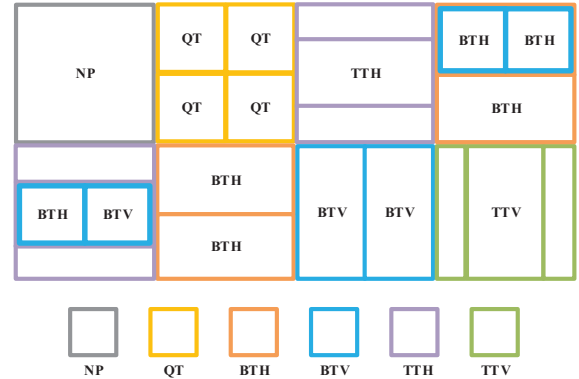}
	\end {minipage}
	\caption{A typical best CU partitions.}
	\label{partition}
\end{figure}

For VVC, intra prediction is  essential for the block-based coding standard. Many new techniques  have been integrated into intra prediction of VVC, including the Quadtree with Nested Multi-Type Tree (QTMT) \cite{introduceVVC2}, 67 angular modes, Multiple-Reference Line (MRL) \cite{introduction-JVET-J0065}, Intra Sub-Partitions (ISP) \cite{introduction-icip2019} and Intra Block Copy (IBC) \cite{introduction-jvet-j0042},   and so forth. These techniques lead to a higher coding complexity, especially  for QTMT. The paper \cite{introdcution-Tissier19} has verified that if the unnecessary splits were not tested, the encoding time could  save 97.5\% on VTM3.0 under the All Intra (AI) configuration. Therefore, it is imperative to decrease the complexity of VVC while keeping the coding efficiency.  It is worthy to notice that the straightforward scheme of the complexity reduction is to terminate the Coding Unit (CU) partition after testing
the Non-Partition (NP) mode or skip some unnecessary split modes  (\textit{e.g.},  Quadtree (QT), Horizontal Binary Tree (BTV), Vertical Binary Tree (BTV), Horizontal Ternary Tree (TTH), or Vertical Ternary Tree (TTV), as illustrated in Fig. \ref{partition}).

To date, the predominant complexity optimization methods are to predict the necessary split modes by data-driven approaches \cite{ datadriven-Zhu18,statistical-Yang20,datadriven-Park21,datadriven-liTIP21, Fu2018} and  statistical approaches \cite{STATISTICAL-Mai05,statistical-Fu19,STATISTICAL-Saldanha20, Tang2018,Gu2018}. For data-driven approaches,  Yang \textit{et al.} \cite{statistical-Yang20} proposed a learning-based fast CU partition approach for intra-prediction. The textural complexity of the current CU, local texture information and the context information of neighboring CUs are used as the features to train the decision tree classifiers. Park \textit{et al.} \cite{datadriven-Park21} fed the extracted features (\textit{e.g.}, QT depth, BT/TT depth, block shape ratio) into the designed Lightweight Neural Network (LNN). Based on the trained model,  the method can optimize the complexity by terminating the  TTH and TTV split modes.  Li \textit{et al.} \cite{datadriven-liTIP21} established a large-scale database and proposed a multi-stage exit CNN model to determine the partition modes. This learning method is adapted to different CU sizes for  complexity optimization.  For statistical approaches, Mai \textit{et  al.} \cite{STATISTICAL-Mai05} proposed Structural Similarity Index Metric (SSIM \cite{SSIM-wang04})-based fast decision methods, in which Rate Distortions (RD) was utilized to determine the split modes of CUs.  Fu \textit{et  al.} \cite{statistical-Fu19} optimized the complexity by two steps. Firstly, the RD cost of horizontal binary-tree mode is checked. Then, a Bayesian-based classifier is used to decide whether to skip other split modes. In \cite{STATISTICAL-Saldanha20}, the Fast Decision Based on Variance (FDV) and Fast Decision Based on Intra Sub-Partition  (FD-ISP) strategies are designed to decide whether to skip the binary and ternary partition. In summary,  most of the above methods only consider the features (\textit{e.g.}, mean, variance, RD cost and gradient) of the neighbor CUs and the  current CU. However, the relationship between the split mode selection  and the sub CUs characteristic  of the CU has not been investigated.

To solve the above  problem, we propose a Fast
Decision Scheme Based on Structural Similarity Index Metric
Variation (FDS-SSIMV) in this letter. The contributions of this letter are summarized as follows: 
\begin{itemize}
	\item A SSIM Variation (SSIMV) measure scheme is proposed to evaluate the asymmetry characteristic among the sub CUs of the corresponding split modes. The asymmetry characteristic is first observed and then evaluated with SSIM.
	\item The FDS-SSIMV  is designed to select the indispensable split modes and finally optimize coding complexity. The SSIMV value of the split mode is designed to indicate  the necessary split modes. Additionally, we design the SSIMV of the QT split mode as a role to adjust the number of  skipped split modes. 
	\item   Experimental results demonstrate  that the proposed method can reduce 64.74\% average Time Saving (TS) with negligible method overhead. In addition, the performance of the method outperforms the compared methods.
\end{itemize}

\section{Fast CU Partition Scheme}

\subsection{Review and Motivation}
\subsubsection{Review of QTMT Partition}
In VVC, the QTMT partition structure has been  introduced to enhance the coding performance. The Coding Tree Unit (CTU) firstly performs QT partition as the root node, and then the leaf nodes are set to test all permitted partition modes (\textit{e.g.}, NP, QT, BTH, BTV, TTH, or TTV, as illustrated in Fig. \ref{partition}). After recursive coding, the smallest RD  cost of the CU can be obtained. It can be expressed as:
\begin{equation}
\begin{aligned}
J_{CU}=\min_{m\in M } \left \{ D_{m}+\lambda\times R_{m} \right \} 
\end{aligned},
\label{minmum_RDcost}
\end{equation}
where $J_{CU}$ is the minimum RD cost of the current CU. $m$ and $M$ are the index of the current test partition mode and the number of all permitted partition modes for the current CU. $D_{m}$ and $R_{m}$ are  distortion and  bit rate of the $m$-th partition  mode of the CU, respectively. $\lambda$ is the Lagrangian parameter.  As illustrated above, the partition mode with the smallest RD cost is the best partition scheme. Therefore, the complexity optimization method can accelerate the video coding by skipping the split modes with a higher RD cost to accelerate the coding effectively. 

\begin{figure}[t]

	\begin {minipage} [htbp] {0.42\textwidth}
	\centering
	\includegraphics [width=1.0\textwidth] {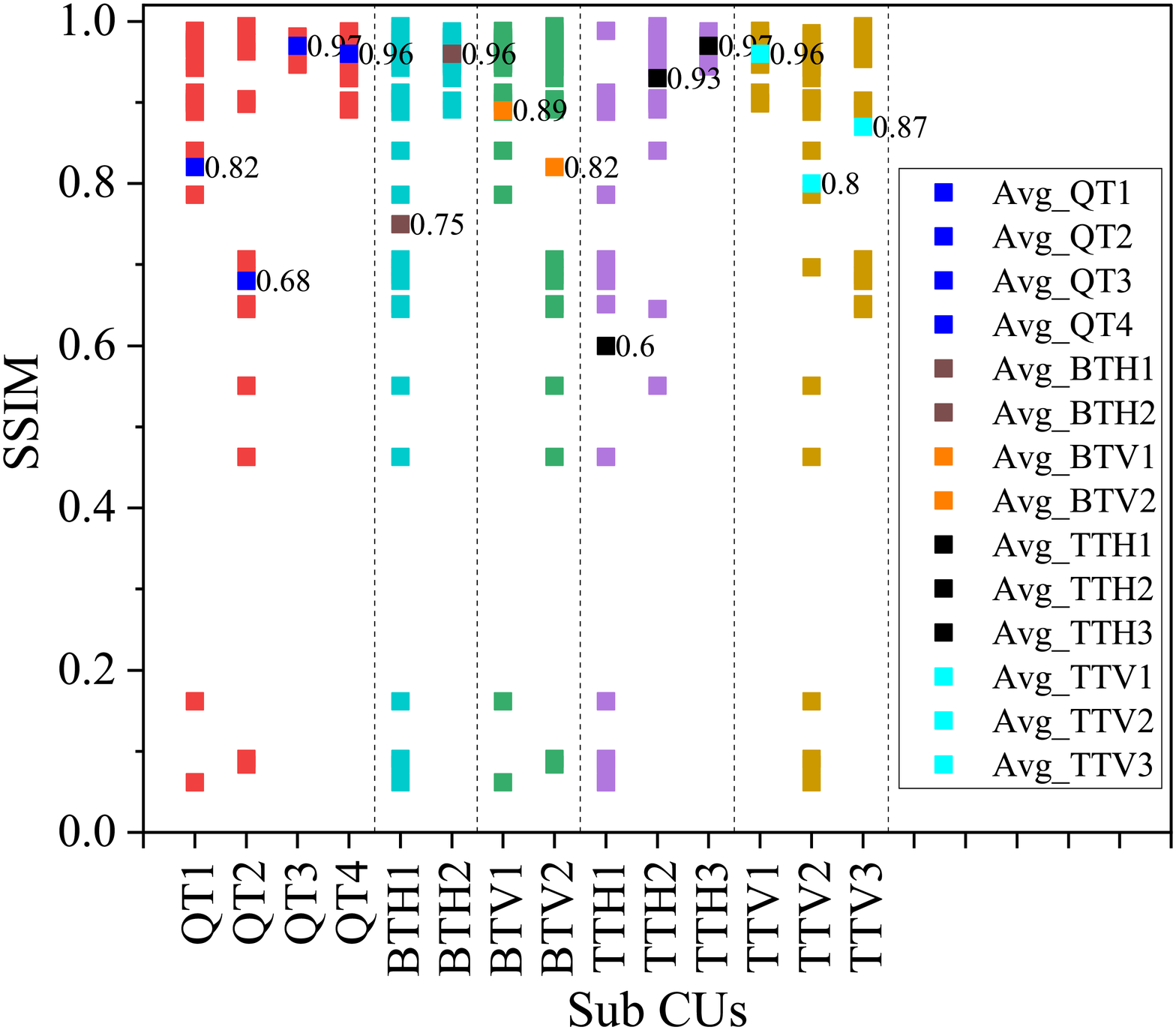}
	\end {minipage}
	\caption{ The SSIM values of sub CUs for  different split modes.}
	\label{ssimdshceme3}
\end{figure}

\subsubsection{Motivation}
According to Eq. (\ref{minmum_RDcost}), the complexity optimization method needs to decide the most necessary split mode to be tested by terminating  the partition or skipping some split modes.  However, as shown in Fig. \ref{partition}, the best partition  mode of parent CU has many different  sub best partition modes. It implies that the CU has the asymmetry characteristic under the split modes.  

To evaluate the asymmetry of the CU, the SSIM metric \cite{SSIM-wang04} is adapted. This experiment is carried out  on the 32$\times$32 CU. The typical 32$\times$32 CU is divided into sixty-four 4$\times$4 non-overlapped 
blocks. The SSIM values of these non-overlapped 
blocks are calculated with the corresponding original CU blocks and predicted CU blocks. The detailed formula is listed as:   
\begin{equation}
\begin{aligned}
SSIM(o,p)=\frac{(2\mu_{o}\mu_{p}+c_{1})(2\sigma _{o}\sigma_{p}+c_{2}) }{(\mu_{o}^2 +\mu_{p}^2+c_{1})(\sigma_{o}^2 +\sigma_{p}^2+c_{1})
}
\end{aligned},
\end{equation}
where $o$ and $p$ are the original CU and predicted CU with NP mode, respectively. $\mu_{o}$ and $ \mu_{p}$ are the average values of the original  CU and predicted CU. $\sigma_{o}$ and $ \sigma_{p}$ are the variance values of the original CU and predicted CU. $c_{1}$ and $ c_{2}$ are the constant values. They are set as $c_{1}=(k_{1}L)^2$ and $c_{2}=(k_{2}L)^2$. $k_{1}$ is set as 0.01. $k_{2}$ is set as 0.03. $L$ is set as $2^{10}-1$. The size of the Gaussian filter is set as 4$\times$4.

After that, according to the split modes, SSIM values of non-overlapping blocks are classified into  different categories (QT1, QT2, QT3 and QT4 for QT split mode, BTH1 and BTH2 for BTH split mode, BTV1 and BTV2 for BTV split mode, TTH1, TTH2 and TTH3 for TTH split mode and TTV1, TTV2 and TTV3 for TTV split mode). As illustrated  in Fig. \ref{ssimdshceme3}, it indicates  the SSIMV characteristic of the CU. Especially for TTH split mode, the average SSIM values of the sub CUs show a large variation. This characteristic indicates TTH1 sub CU of  this split mode has a significant  difference between the original CU and predicted CU, and other sub CUs have a higher similarity than the other split mode. Therefore, it indicates this split mode can split the CU more precisely and have a more chance to obtain the smallest RD cost. Then, the  degree of the asymmetry of the CU under different split modes may indicate the necessary split modes of the CU.

\subsection{SSIM-Variation Calculation Scheme}
\label{ssim-feature}

Based on the above analysis, we designed the calculation schemes to  evaluate the  asymmetry of the CU for different split modes. The notations of the SSIMV value for the split modes are expressed as $V_{\rm QT}$, $V_{\rm BTH}$, $V_{\rm BTV}$, $V_{\rm TTH}$ and $V_{\rm TTV}$, respectively. The SSIMV values of these partition modes are calculated as follows. 

\begin{figure}[t]
	\centering
	\begin {minipage} [htbp] {0.11\textwidth}
	\centering
	\includegraphics [width=1.0\textwidth] {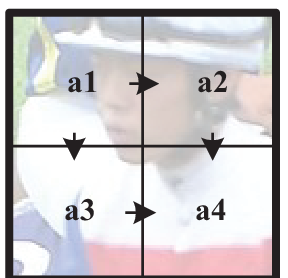}
	\centerline{(a) QT}
	\end {minipage}
	\begin {minipage} [htbp] {0.11\textwidth}
	\centering
	\includegraphics [width=1.0\textwidth] {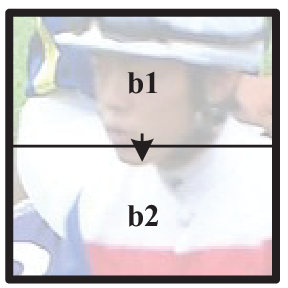}
	\centerline{(b) BTH}
	\end {minipage}
	\begin {minipage} [htbp] {0.11\textwidth}
	\centering
	\includegraphics [width=1.0\textwidth] {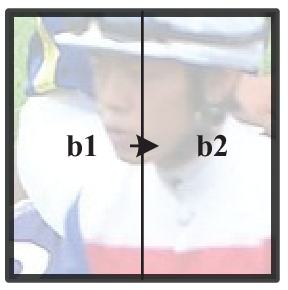}
	\centerline{(c) BTV}
	\end {minipage}
	\begin {minipage} [htbp] {0.345\textwidth}
	\centering
	\includegraphics [width=1.0\textwidth] {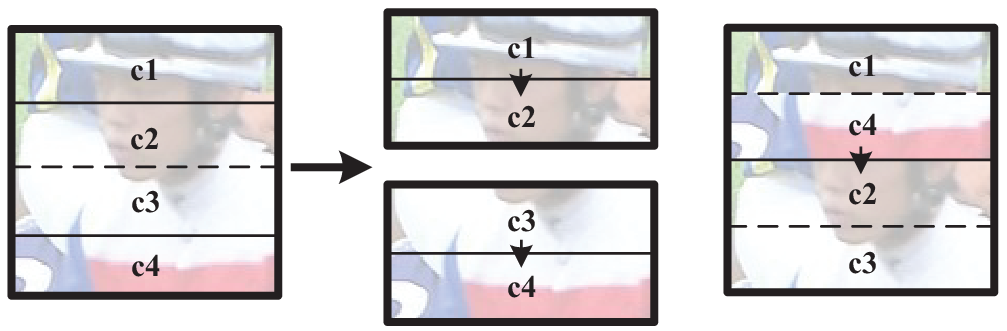}
	\centerline{(d) TTH}
	\end {minipage}
	\begin {minipage} [htbp] {0.345\textwidth}
	\centering
	\includegraphics [width=1.0\textwidth] {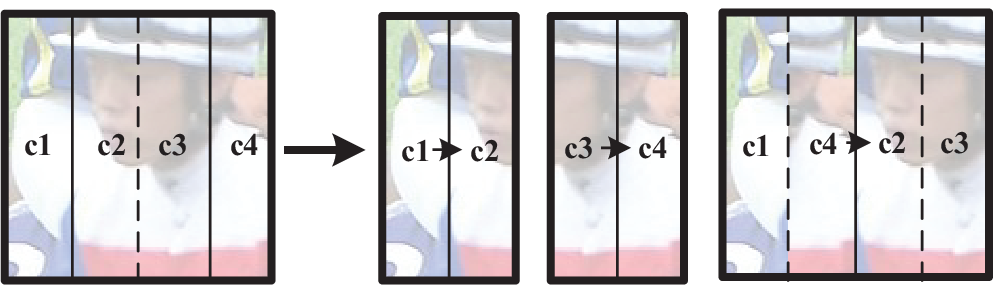}
	\centerline{(e) TTV}
	\end {minipage}				
	\caption{ The scheme of the SSIMV calculation for the typical CU.}
	\label{ssimdshceme}
\end{figure}

Firstly,  $V_{\rm QT}$ can be obtained by:
\begin{equation}
\begin{aligned}
&V_{\rm QT}=\frac{D_{\rm A1}+D_{\rm A2}+D_{\rm A3}+D_{\rm A4}}{4},	
\end{aligned}
\end{equation}
where $D_{\rm A1}$, $D_{\rm A2}$, $D_{\rm A3}$ and $D_{\rm A4}$ are the difference in the SSIM values of the neighbor sub CUs. They are expressed as:

\begin{equation}
\label{eq:kkt}
\begin{aligned}
\left\{ {\begin{array}{ll}
	\vspace{4pt}
	{D_{\rm A1}=\left |S_{\rm a1}-S_{\rm a2}  \right |}  \\
	\vspace{4pt}
	{D_{\rm A2}=\left |S_{\rm a1}-S_{\rm a3}\right |}   \\
	\vspace{4pt}
	{D_{\rm A3}=\left |S_{\rm a2}-S_{\rm a4}\right | }\\
		\vspace{4pt}
		{D_{\rm A4}=\left |S_{\rm a3}-S_{\rm a4}\right |}   \\
	\end{array}
} 
\right .,
\end{aligned}
\end{equation}
where $S_{\rm a1}$, $S_{\rm a2}$, $S_{\rm a3}$ and $S_{\rm a4}$ are the SSIM values of the corresponding  sub CUs in Fig. \ref{ssimdshceme}(a), respectively. The size of the Gaussian filter used for SSIM calculation is set as 11$\times$11 for the size of sub-CUs greater than 11. Otherwise, it is set as 4$\times$4.

Similarly,  $V_{\rm BTH}$ and $V_{\rm BTV}$ can also be obtained by:
\begin{equation}
\small
\begin{aligned}
&V_{\rm BTH}=V_{\rm BTV}=\left |S_{\rm b1}-S_{\rm b2}\right |,	
\end{aligned}
\end{equation}
where $S_{\rm b1}$ and $S_{\rm b2}$ are the SSIM values of the corresponding sub CUs in Fig. \ref{ssimdshceme}(b, c), respectively.

Besides,   $V_{\rm TTH}$ and $V_{\rm TTV}$ can be calculated by:
\begin{equation}
\small
\begin{aligned}
&V_{\rm TTH}=V_{\rm TTV}=\frac{D_{\rm C1}+D_{\rm C2}+D_{\rm C3}}{4}		
\end{aligned}.
\end{equation} 
where $D_{\rm C1}$, $D_{\rm C2}$ and $D_{\rm C3}$ are the difference in the SSIM values of neighbor sub CUs. The sub CUs  of the TTH and TTV split modes, which are used to evaluated the SSIMV value,  are designed as Fig. \ref{ssimdshceme}(d, e). $D_{\rm C1}$, $D_{\rm C2}$ and  $D_{\rm C3}$ are expressed as:
\begin{equation}
\label{eq:kkt}
\begin{aligned}
\left\{ {\begin{array}{ll}
	\vspace{4pt}
	{D_{\rm C1}=\left |S_{\rm c1}-S_{\rm c2}  \right |}  \\
	\vspace{4pt}
	{D_{\rm C2}=\left |S_{\rm c1}+S_{\rm c4}-S_{\rm c2}-S_{\rm c3}\right |}   \\
	{D_{\rm C3}=\left |S_{\rm c3}-S_{\rm c4}\right | }
	\end{array}
} 
\right .,
\end{aligned}
\end{equation}
where $S_{\rm c1}$, $S_{\rm c2}$, $S_{\rm c3}$ and $S_{\rm c4}$ are the SSIM values of the corresponding sub CUs in Fig. \ref{ssimdshceme}(d, e), respectively. 
\begin{figure}[t]
	\small
	\centering
	\begin {minipage} [htbp] {0.43\textwidth}
	\centering
	\includegraphics [width=1\textwidth] {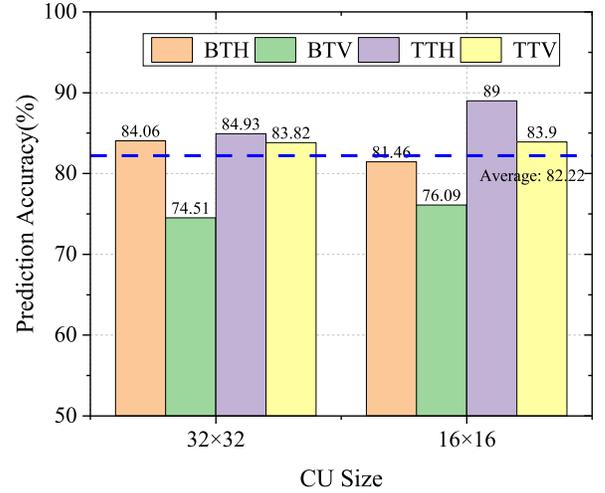}
	\end {minipage}
	\caption{The prediction accuracy of FDS-SSIMV.}
	\label{predictionaccuracy3}
\end{figure} 
\begin{figure}[t]
	\small
	\centering
	\begin {minipage} [htbp] {0.4\textwidth}
	\centering
	\includegraphics [width=1\textwidth] {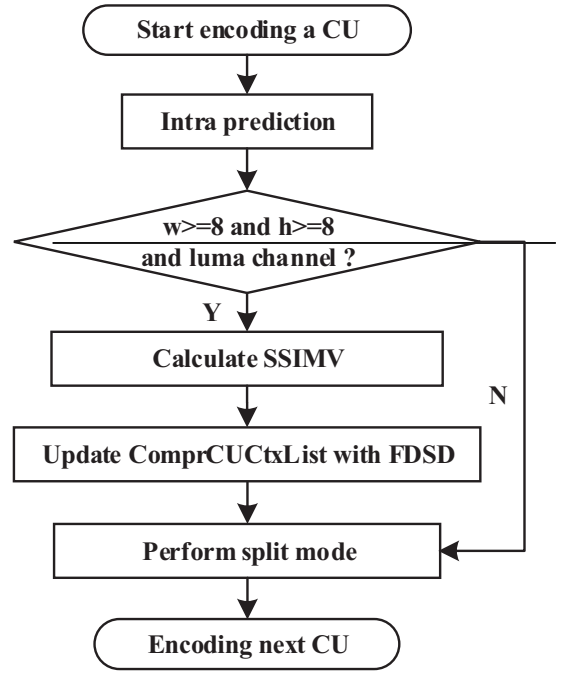}
	\end {minipage}
	\caption{The flowchart of the proposed method.}
	\label{flowchart}
\end{figure}

\begin{table*}[t]
		\caption{Performance of  the benchmarks and proposed method on VTM 7.0 } \label{tab:BDBR}
		\centering
			\scalebox{0.85}{
\renewcommand{\arraystretch}{1.15}
\begin{tabular}{|cc|cccc|cccc|c|}
	\hline
	\multicolumn{2}{|c|}{\textbf{Sequences}}                                       & \multicolumn{4}{c|}{\textbf{TS (\%)}}                                                                                                                                           & \multicolumn{4}{c|}{\textbf{BDBR (\%)}}                                                                                                                                     & \textbf{Overhead \textbf{(\%)}}                      \\ \hline
	\multicolumn{1}{|c|}{\textbf{Class}}                & \textbf{Name}            & \multicolumn{1}{c|}{\textbf{Fu-ICME'19}} & \multicolumn{1}{c|}{\textbf{Yang-TCSVT'20}} & \multicolumn{1}{c|}{\textbf{Li-TIP'21}}              & \textbf{Proposed}                     & \multicolumn{1}{c|}{\textbf{Fu-ICME'19}} & \multicolumn{1}{c|}{\textbf{Yang-TCSVT'20}}                    & \multicolumn{1}{c|}{\textbf{Li-TIP'21}} & \textbf{Proposed}           & \multicolumn{1}{c|}{\textbf{$OH_{pro}$}} \\ \hline
\multicolumn{1}{|c|}{}                              & \textbf{Tango2}          & \multicolumn{1}{c|}{52.47}            & \multicolumn{1}{c|}{45.28}             & \multicolumn{1}{c|}{54.35}                        & 61.71                                 & \multicolumn{1}{c|}{3.37}             & \multicolumn{1}{c|}{1.63}                                 & \multicolumn{1}{c|}{3.49}            & 2.24                        & 1.08                                   \\ \cline{2-11} 
\multicolumn{1}{|c|}{}                              & \textbf{FoodMarkrt4}     & \multicolumn{1}{c|}{53.33}            & \multicolumn{1}{c|}{60.49}             & \multicolumn{1}{c|}{57.91}                        & 69.42                                 & \multicolumn{1}{c|}{2.35}             & \multicolumn{1}{c|}{5.15}                                 & \multicolumn{1}{c|}{2.78}            & 2.48                        & 0.64                                   \\ \cline{2-11} 
\multicolumn{1}{|c|}{\multirow{-3}{*}{\textbf{A1}}} & \textbf{Campfire}        & \multicolumn{1}{c|}{61.68}            & \multicolumn{1}{c|}{50.12}             & \multicolumn{1}{c|}{66.52}                        & 61.87                                 & \multicolumn{1}{c|}{4.33}             & \multicolumn{1}{c|}{2.64}                                 & \multicolumn{1}{c|}{4.17}            & 2.71                        & 0.48                                   \\ \hline
\multicolumn{1}{|c|}{}                              & \textbf{Catrobot1}       & \multicolumn{1}{c|}{59.75}            & \multicolumn{1}{c|}{47.17}             & \multicolumn{1}{c|}{62.87}                        & 63.73                                 & \multicolumn{1}{c|}{6.75}             & \multicolumn{1}{c|}{1.13}                                 & \multicolumn{1}{c|}{4.88}            & 3.10                        & 0.64                                   \\ \cline{2-11} 
\multicolumn{1}{|c|}{}                              & \textbf{DaylightRoad2}   & \multicolumn{1}{c|}{60.83}            & \multicolumn{1}{c|}{52.44}             & \multicolumn{1}{c|}{65.63}                        & 69.42                                 & \multicolumn{1}{c|}{2.80}             & \multicolumn{1}{c|}{2.18}                                 & \multicolumn{1}{c|}{2.78}            & 2.48                        & 0.80                                   \\ \cline{2-11} 
\multicolumn{1}{|c|}{\multirow{-3}{*}{\textbf{A2}}} & \textbf{ParkRunning3}    & \multicolumn{1}{c|}{62.90}            & \multicolumn{1}{c|}{45.58}             & \multicolumn{1}{c|}{63.01}                        & 54.82                                 & \multicolumn{1}{c|}{2.69}             & \multicolumn{1}{c|}{1.25}                                 & \multicolumn{1}{c|}{2.68}            & 1.30                        & 0.38                                   \\ \hline
\multicolumn{1}{|c|}{}                              & \textbf{MarketPlace}     & \multicolumn{1}{c|}{59.98}            & \multicolumn{1}{c|}{57.97}             & \multicolumn{1}{c|}{65.15}                        & 70.61                                 & \multicolumn{1}{c|}{2.00}             & \multicolumn{1}{c|}{4.20}                                 & \multicolumn{1}{c|}{1.89}            & 2.06                        & 0.69                                   \\ \cline{2-11} 
\multicolumn{1}{|c|}{}                              & \textbf{RitualDance}     & \multicolumn{1}{c|}{57.17}            & \multicolumn{1}{c|}{61.93}             & \multicolumn{1}{c|}{62.98}                        & 68.23                                 & \multicolumn{1}{c|}{3.86}             & \multicolumn{1}{c|}{3.73}                                 & \multicolumn{1}{c|}{2.69}            & 3.16                        & 0.80                                   \\ \cline{2-11} 
\multicolumn{1}{|c|}{}                              & \textbf{BasketballDrive} & \multicolumn{1}{c|}{58.95}            & \multicolumn{1}{c|}{55.07}             & \multicolumn{1}{c|}{68.50}                        & 67.04                                 & \multicolumn{1}{c|}{3.55}             & \multicolumn{1}{c|}{2.19}                                 & \multicolumn{1}{c|}{3.87}            & 3.44                        & 0.70                                   \\ \cline{2-11} 
\multicolumn{1}{|c|}{}                              & \textbf{BQTerrance}      & \multicolumn{1}{c|}{55.42}            & \multicolumn{1}{c|}{58.96}             & \multicolumn{1}{c|}{64.39}                        & 65.76                                 & \multicolumn{1}{c|}{1.75}             & \multicolumn{1}{c|}{5.25}                                 & \multicolumn{1}{c|}{2.57}            & 2.53                        & 0.62                                   \\ \cline{2-11} 
\multicolumn{1}{|c|}{\multirow{-5}{*}{\textbf{B}}}  & \textbf{Cactus}          & \multicolumn{1}{c|}{59.55}            & \multicolumn{1}{c|}{60.68}             & \multicolumn{1}{c|}{67.70}                        & 67.34                                 & \multicolumn{1}{c|}{3.54}             & \multicolumn{1}{c|}{1.31}                                 & \multicolumn{1}{c|}{2.85}            & 3.05                        & 0.56                                   \\ \hline
\multicolumn{1}{|c|}{}                              & \textbf{BasketballDrill} & \multicolumn{1}{c|}{58.39}            & \multicolumn{1}{c|}{60.09}             & \multicolumn{1}{c|}{60.98}                        & 63.63                                 & \multicolumn{1}{c|}{4.29}             & \multicolumn{1}{c|}{4.29}                                 & \multicolumn{1}{c|}{4.72}            & 4.49                        & 0.51                                   \\ \cline{2-11} 
\multicolumn{1}{|c|}{}                              & \textbf{BQMall}          & \multicolumn{1}{c|}{58.91}            & \multicolumn{1}{c|}{55.04}             & \multicolumn{1}{c|}{67.45}                        & 65.93                                 & \multicolumn{1}{c|}{4.19}             & \multicolumn{1}{c|}{2.84}                                 & \multicolumn{1}{c|}{3.10}            & 3.58                        & 0.49                                   \\ \cline{2-11} 
\multicolumn{1}{|c|}{}                              & \textbf{PartyScene}      & \multicolumn{1}{c|}{57.67}            & \multicolumn{1}{c|}{57.20}             & \multicolumn{1}{c|}{64.64}                        & 63.68                                 & \multicolumn{1}{c|}{1.94}             & \multicolumn{1}{c|}{2.78}                                 & \multicolumn{1}{c|}{1.86}            & 1.92                        & 0.38                                   \\ \cline{2-11} 
\multicolumn{1}{|c|}{\multirow{-4}{*}{\textbf{C}}}  & \textbf{RaceHourses}     & \multicolumn{1}{c|}{58.90}            & \multicolumn{1}{c|}{54.91}             & \multicolumn{1}{c|}{65.68}                        & 64.95                                 & \multicolumn{1}{c|}{3.18}             & \multicolumn{1}{c|}{2.39}                                 & \multicolumn{1}{c|}{2.50}            & 2.20                        & 0.37                                   \\ \hline
\multicolumn{1}{|c|}{}                              & \textbf{BasketballPass}  & \multicolumn{1}{c|}{57.77}            & \multicolumn{1}{c|}{54.69}             & \multicolumn{1}{c|}{62.62}                        & 62.28                                 & \multicolumn{1}{c|}{3.38}             & \multicolumn{1}{c|}{1.88}                                 & \multicolumn{1}{c|}{3.66}            & 2.92                        & 0.42                                   \\ \cline{2-11} 
\multicolumn{1}{|c|}{}                              & \textbf{BlowingBubbles}  & \multicolumn{1}{c|}{55.27}            & \multicolumn{1}{c|}{52.42}             & \multicolumn{1}{c|}{61.85}                        & 61.72                                 & \multicolumn{1}{c|}{2.28}             & \multicolumn{1}{c|}{3.17}                                 & \multicolumn{1}{c|}{2.38}            & 2.03                        & 0.32                                   \\ \cline{2-11} 
\multicolumn{1}{|c|}{}                              & \textbf{BQSquare}        & \multicolumn{1}{c|}{54.17}            & \multicolumn{1}{c|}{51.36}             & \multicolumn{1}{c|}{62.52}                        & 65.47                                 & \multicolumn{1}{c|}{1.13}             & \multicolumn{1}{c|}{1.37}                                 & \multicolumn{1}{c|}{2.04}            & 1.69                        & 0.40                                   \\ \cline{2-11} 
\multicolumn{1}{|c|}{\multirow{-4}{*}{\textbf{D}}}  & \textbf{RaceHorses}      & \multicolumn{1}{c|}{56.62}            & \multicolumn{1}{c|}{45.47}             & \multicolumn{1}{c|}{61.29}                        & 62.21                                 & \multicolumn{1}{c|}{3.42}             & \multicolumn{1}{c|}{1.19}                                 & \multicolumn{1}{c|}{2.92}            & 2.08                        & 0.37                                   \\ \hline
\multicolumn{1}{|c|}{}                              & \textbf{FourPeople}      & \multicolumn{1}{c|}{57.87}            & \multicolumn{1}{c|}{51.51}             & \multicolumn{1}{c|}{66.91}                        & 68.82                                 & \multicolumn{1}{c|}{3.77}             & \multicolumn{1}{c|}{1.66}                                 & \multicolumn{1}{c|}{3.30}            & 4.37                        & 0.71                                   \\ \cline{2-11} 
\multicolumn{1}{|c|}{}                              & \textbf{Johnny}          & \multicolumn{1}{c|}{57.79}            & \multicolumn{1}{c|}{57.49}             & \multicolumn{1}{c|}{64.35}                        & 67.14                                 & \multicolumn{1}{c|}{6.48}             & \multicolumn{1}{c|}{2.42}                                 & \multicolumn{1}{c|}{5.08}            & 4.12                        & 0.74                                   \\ \cline{2-11} 
\multicolumn{1}{|c|}{\multirow{-3}{*}{\textbf{E}}}  & \textbf{KristenAndSara}  & \multicolumn{1}{c|}{56.17}            & \multicolumn{1}{c|}{58.85}             & \multicolumn{1}{c|}{66.11}                        & 65.50                                 & \multicolumn{1}{c|}{4.71}             & \multicolumn{1}{c|}{3.78}                                 & \multicolumn{1}{c|}{3.93}            & 3.66                        & 0.70                                   \\ \hline
\multicolumn{2}{|c|}{\textbf{ALL Average}}                                     & \multicolumn{1}{c|}{57.80}            & \multicolumn{1}{c|}{54.32}             & \multicolumn{1}{c|}{{\color[HTML]{3531FF} 64.17}} & {\color[HTML]{FF0000} \textbf{64.74}} & \multicolumn{1}{c|}{3.44}             & \multicolumn{1}{c|}{{\color[HTML]{FF0000} \textbf{2.66}}} & \multicolumn{1}{c|}{3.29}            & {\color[HTML]{3531FF} 2.79} & 0.60                                   \\ \hline
\end{tabular}}
\end{table*}
Based on the SSIMV feature, we propose a scheme called FDS-SSIMV. The scheme  can select  the necessary test split modes in VVC. Additionally, the SSIMV characteristic of the split modes is used to evaluate the asymmetry of the CU. The higher SSIMV value of the split mode  indicates that the sub CUs of the spit mode have a more significant  variation. To a certain extent, the phenomenon implies  that the necessary  split mode of the CU has a higher  SSIMV value.  Therefore, we sort  the SSIMV values of all the split modes to determine the most needed testing split mode. The number of skipping split modes is set as half of the permitted number of the CU to guarantee the coding quality. It is noteworthy that skipping the QT partition may lead to a huge loss in encoding performance. Therefore, our fast algorithm has been designed not to skip the QT split mode. This design adjusts the number of  skipped split modes to balance the coding quality and efficiency.

To further verify the FDS-SSIMV performance, the  scheme is implemented in the original VVC reference software  (VTM7.0) under the AI configuration and Quantization Parameters (QPs) 22,27, 32 and 37. The average prediction accuracy of the 32$\times$32 and 16$\times$16 CU sizes  are tested on   the sequences (\textit{Tango2, ParkRunning3, BQTerrace, PartyScene, BQSquare and KristenAndSara}).  As illustrated in Fig. \ref{predictionaccuracy3}, it is implied that  the FDS-SSIMV  can obtain the high average prediction accuracy with 82.22\% for BTH, BTV, TTH and TTV in the 32$\times$32 and 16$\times$16 CU sizes.

Combined with the above analysis,  the flowchart of the proposed method  is shown in Fig. \ref{flowchart}. In this method, the optimization is designed for the CU size larger than or equal to 8$\times$8 CU and the luma channel of the CU. Then, the SSIMV values of the permitted  split modes are calculated.  After that, the testing stack (ComprCUCtxList) is  updated with  FDS-SSIMV. Based on this updated ComprCUCtxList, the requisite split modes are tested as the original reference software.
\section{Expertiment Results}

 The proposed method is integrated  into VTM7.0 to verify its effectiveness. This experiment is tested on the mandatory video sequences of the Common Testing Coding (CTC) JVET-N1010 \cite{CTC-F.Bossen20} under the AI configuration and QPs 22, 27, 32 and 37.  Furthermore, the performances of the deep-learning-based and statistical complexity optimization methods \cite{statistical-Yang20,datadriven-liTIP21, statistical-Fu19} are also compared in this section. There are denoted as Yang-TCSVT'20, Li-TIP'21 and Fu-ICME'19, respectively.

In this section, the BDBR and TS performance are listed in Table \ref{tab:BDBR}. The BDBR metric shows a negative correlation with the RD performance of the method. 
In Table \ref{tab:BDBR},   the average  BDBR of our method is 2.79\%. The BDBR performance of the proposed method outperforms the Fu-ICME'19 and Li-TIP'21 methods by 0.65\% and 0.5\%, respectively. Although the BDBR of the Yang-TCSVT'20 method is competitive  with the proposed method, the average TS is lower than the proposed method by 10.42\%.  For TS performance, it indicates the degree of complexity reduction of the proposed method. It is calculated by $
TS=\left | T_{org}-T_{pro} \right | /T_{org} \times 100\%$. $T_{org}$ and $T_{pro}$ are the total time cost of the sequence encoding with the reference software and the integrated software, respectively. The proposed method reduces 54.82\%$\sim$70.61\% TS. The average TS of the proposed method is 64.74\%, which is superior to the average TS of the Fu-ICME'19, Yang-TCSVT'20,  and Li-TIP'21 methods with 57.8\%, 54.32\% and 64.17\%, respectively. 

To evaluate the method's performance, we also calculate the overhead of the proposed method in the original reference software and the integrated software with our method. The overhead are defined by $OH_{pro}$, which can be obtained as $OH_{org}=T_{ssim} /T_{pro} \times 100\% $.  $T_{ssim}$ is the time cost of the proposed method in running the integrated software to coding the sequences.  In Table \ref{tab:BDBR}, the average overhead of the $OH_{pro}$ is 0.6\%. The overhead of the proposed method is decreased than the 3.67\% overhead of the Li-TIP'21 method \cite{datadriven-liTIP21} with the aid of 3.07\%. In other words, the proposed method with negligible overhead is a light scheme.

In summary, the proposed method achieves an average TS of 64.74\%, an average BDBR of 2.79\% and an average overhead of 0.60\%, which outperforms the compared methods and  indicates the efficiency of the scheme.

\section{Conclusion}
In this letter, we first observe the asymmetry characteristic of the  CUs. Then, the SSIMV calculation schemes  for different split modes are designed to evaluate the asymmetry characteristic. Based on this, the FDS-SSIMV  is proposed in this letter. The experimental results show that the method has  low overhead, high TS and competitive  BDBR. Compared with the benchmarks, the performance of the proposed method is superior to them.

\bibliographystyle{IEEEbib}
\bibliography{icme2022template}

\end{document}